\documentclass[11pt]{article}
\usepackage[utf8]{inputenc}
\usepackage{amsfonts}
\usepackage{amsmath}
\usepackage{amssymb}
\usepackage{indentfirst}
\usepackage{graphicx}
\usepackage[colorlinks]{hyperref}
\usepackage{cite}
\usepackage{colortbl}
\usepackage{soul}

\usepackage{microtype}
\usepackage[normalem]{ulem}

\numberwithin{equation}{section}
\allowdisplaybreaks

\begin{document}

\baselineskip=18pt %a la harvmac
\baselineskip 0.6cm
\begin{titlepage}
\vskip 4cm

\begin{center}
\textbf{\LARGE Enlarged super-$\mathfrak{bms}_{3}$ algebra and its flat limit}
\par\end{center}{\LARGE \par}

\begin{center}
	\vspace{1cm}
    \textbf{Javier Matulich}$^{\ast}$,
	\textbf{Evelyn Rodríguez}$^{\dag, \star}$
	\small
	\\[5mm]
	$^{\ast}$\textit{Instituto de F\'isica Te\'orica UAM/CSIC,}\\
	\textit{C/ Nicol\'as Cabrera 13-15, Universidad Aut\'onoma de Madrid, Cantoblanco, Madrid 28049,Spain. }
  \\[2mm]
 $^{\star}$\textit{Grupo de Investigación en F\'isica Te\'orica, GIFT, }\\
	\textit{Concepci\'on, Chile.}
        \\[2mm]
    $^{\dag}$\textit{Departamento de Matem\'atica y F\'isica Aplicadas, }\\
	\textit{ Universidad Cat\'olica de la Sant\'isima Concepci\'on, }\\
\textit{ Alonso de Ribera 2850, Concepci\'on, Chile.}
 \\[5mm]
	\footnotesize
	\texttt{erodriguez@ucsc.cl},
    \texttt{javier.matulich@csic.es}
	\par\end{center}
\vskip 26pt
%\begin{abstract}
\centerline{{\bf Abstract}}
\medskip
\noindent  
In this paper we analyze the asymptotic symmetries of the three-dimensional Chern-Simons supergravity for a  supersymmetric extension of the semi-simple enlargement of the Poincaré algebra, also known as AdS-Lorentz superalgebra, which is characterized by two fermionic generators. We propose a consistent set of asymptotic boundary conditions for the aforementioned supergravity theory, and we show that the corresponding charge algebra defines a supersymmetric extension of the semi-simple enlargement of the $\mathfrak{bms}_{3}$ algebra, with three independent central charges. This asymptotic symmetry algebra can alternatively be written as the direct sum of three copies of the Virasoro algebra, two of which are augmented by supersymmetry. Interestingly, we show that the flat limit of the obtained asymptotic algebra corresponds to a deformed super-$\mathfrak{bms}_{3}$ algebra, being the charge algebra of the minimal Maxwell supergravity theory in three dimensions. 

%\end{abstract}
\end{titlepage}\newpage {\baselineskip=12pt \tableofcontents{}}

\section{Introduction}\label{sec1}
Three dimensional gravity theories formulated through the Chern-Simons (CS) formalism are considered to be interesting toy models since they had allowed us to better understand different aspects of the gravitational interaction and underlying laws of quantum gravity. Furthermore, they share many properties with higher-dimensional gravity models which, in general, are more difficult to approach. In the context of the AdS/CFT correspondence \cite{Maldacena:1997re}, three dimensional models play an important role because there is a well and detailed understanding both of the bulk theories as well as their two-dimensional dual \cite{Brown:1986nw, Witten:1988hf, Moore:1989yh, Elitzur:1989nr, Coussaert:1995zp}. In this scenario, the study of three-dimensional Anti-de Sitter (AdS) gravity has received a great interest due to its rich boundary dynamics characterized by the centrally extended conformal algebra in two dimensions \cite{Brown:1986nw}. In the asymptotically flat case, an infinite dimensional asymptotic symmetry algebra can be found at null infinity, given by the $\mathfrak{bms}_3$ algebra \cite{barnich2007classical}. As it is shown in  \cite{Barnich:2012rz, Gonzalez:2013oaa}, both aforesaid asymptotic symmetries are connected through a well-defined flat limit. These results have been extended to incorporate supersymmetry. For AdS supergravity, it was shown that the boundary dynamics is governed by the superconformal algebra \cite{Coussaert:1993jp,Banados:1998pi,Henneaux:1999ib}, while in the flat case, the asymptotic symmetry algebra is given by a supersymmetric extension of the $\mathfrak{bms}_3$ algebra \cite{Barnich:2014cwa,Barnich:2015sca}.

On the other hand, the approach of CS (super)gravity theories based on extensions and deformations of the Poincaré and AdS symmetries has received great attention and  has led to diverse results \cite{Edelstein:2006se,Izaurieta:2009hz,Gonzalez:2014tta, Salgado:2014jka}. These theories include, in general, new gauge fields besides the usual vielbein and spin connection, and  thus the field content is richer compared to General Relativity (GR). For instance, the so-called Maxwell algebra, which contains additional generators and modifies the Poincaré algebra, allows to introduce a new gauge field denoted as the gravitational Maxwell gauge field. The Maxwell algebra was first introduced in four spacetime dimensions to describe a constant Minskowski spacetime in the presence of an electromagnetic background \cite{Schrader:1972zd,Bacry:1970ye,Bacry:1970du,Gomis:2017cmt}. In three space-time dimensions, a gravity theory based on the Maxwell symmetry can be constructed via the CS formalism. This theory turns out to be given by the Einstein- Hilbert term without the cosmological constant plus a non back-reacting matter field coupled to the geometry \cite{Salgado:2014jka}. The Maxwell (super)algebra and its generalizations have been studied in different contexts leading to interesting results \cite{Edelstein:2006se,Izaurieta:2009hz,Concha:2013uhq,Concha:2014vka,Cangemi:1992ri,Duval:2008tr,Hoseinzadeh:2014bla,Caroca:2017izc,Concha:2018zeb,Concha:2018jxx,Concha:2019icz,Chernyavsky:2020fqs,Adami:2020xkm,Caroca:2021bjo,deAzcarraga:2010sw,Durka:2011nf,deAzcarraga:2012qj,Concha:2014tca,Penafiel:2017wfr,Ravera:2018vra,Concha:2018ywv,Salgado-Rebolledo:2019kft,Concha:2019eip,Kibaroglu:2020tbr,Cebecioglu:2021dqb}.

The presence of an additional gauge field in the Maxwell case leads to new effects compared to GR. In particular, it was shown that the Maxwell gauge field modifies not only the vacuum energy and momentum of the stationary configuration but also the asymptotic structure of the three-dimensional Maxwell CS gravity \cite{Concha:2018zeb}. Indeed, it was shown that the asymptotic symmetry of the theory is given by a deformation and extension of the $\mathfrak{bms}_3$ algebra.  On the other hand, Maxwell CS gravity can be obtained as a vanishing cosmological constant limit from the so-called AdS-Lorentz gravity in three-dimensions \cite{Soroka:2006aj, Salgado:2014qqa}, whose underlying symmetry corresponds to a semi-simple extension of the Poincaré algebra. This symmetry allows to introduce a non-vanishing cosmological constant in the CS action, and the stationary solution turns out to be a BTZ-like solution \cite{Concha:2018jjj}.   Furthermore, considering the analysis of asymptotic symmetries, it was also shown in \cite{Concha:2018jjj} that the
deformed $\mathfrak{bms}_3$ algebra, being the asymptotic symmetry of the Maxwell CS gravity, can be derived
as a flat limit of 
the asymptotic symmetry of the AdS-Lorentz CS gravity which is given by a semi-simple enlargement of the $\mathfrak{bms}_3$ symmetry. 

In this work, we extend the asymptotic analysis done in \cite{Concha:2018zeb, Concha:2018jjj} to the AdS-Lorentz supergravity theory in three dimensions, whose underlying symmetry corresponds to a supersymmetric extension of the AdS-Lorentz algebra.  As it is expected, the asymptotic symmetry algebra is given by a supersymmetric extension of the semisimple enlargement of the  $\mathfrak{bms}_3$ algebra, introduced in \cite{Caroca:2019dds}. The motivation for considering this study is twofold. First, it is worth it to analyze how the asymptotic structure is modified by the presence of supersymmetry in the AdS-Lorentz CS gravity theory. Second, although the enlarged super-$\mathfrak{bms}_3$ algebra was first introduced in \cite{Caroca:2019dds} as an expansion of the Virasoro algebra, here we will show, by imposing appropriate boundary conditions, that it indeed corresponds to the asymptotic symmetry of the minimal AdS-Lorentz CS supergravity theory in three space-time dimensions. Furthermore, we show that the  the deformed super-$\mathfrak{bms}_3$ algebra appears  as a vanishing cosmological constant limit of the enlarged super-$\mathfrak{bms}_3$ algebra.

The paper is organized as follows: In section 2, we briefly review the three-dimensional CS supergravity based on the minimal AdS-Lorentz superalgebra and discuss about the flat limit at the level of the CS action, invariant tensors, curvatures and field equations. In section 3, we compute the asymptotic symmetry algebra for the minimal AdS-Lorentz CS supergravit, which results to be an enlargement of the super-$\mathfrak{bms}_3$ algebra.  We provide the suitable fall-off conditions for
the gauge fields at infinity and the gauge transformations preserving the boundary conditions.  We also recall that the obtained infinite-dimensional symmetry algebra can be written as the direct sum of three copies of the Virasoro algebra, two of which are augmented by supersymmetry.  In section 4, we analyze the flat limit, which can be performed throughout all the steps followed in obtaining the asymptotic symmetry algebra for the AdS-Lorentz supergravity.  We show that the vanishing cosmological constant limit $\ell\rightarrow\infty$ of the enlarged super-$\mathfrak{bms}_3$ algebra leads to an infinite-dimensional lift of the minimal Maxwell superalgebra. Section 5 is devoted to discussion and possible future developments.

\section{Minimal AdS-Lorentz Supergravity theory and flat limit}

In this section, we briefly review the so-called minimal AdS-Lorentz supergravity theory first introduced in \cite{Concha:2018jxx}, and whose vanishing cosmological constant limit leads to the minimal supersymmetric extension of the Maxwell gravity \cite{Bonanos:2009wy,Concha:2018jxx}. First, we extend the AdS-Lorentz algebra generated by $\{J_{a},P_{a},Z_{a}\}$ with the
fermionic generators $\{Q_{\alpha },\Sigma _{\alpha }\}$, obtaining the
following superalgebra \footnote{ While there is a supersymmetric extension of the AdS-Lorentz with a single fermionic generator \cite{Fierro:2014lka}, in the flat limit, the four-momentum generators $P_a$ are no longer expressed as bilinear expressions of the fermionic generators $Q$. As a consequence, the action constructed from the contracted superalgebra is an exotic CS supersymmetric action.}
\begin{eqnarray}
\left[ J_{a},J_{b}\right] &=&\epsilon _{abc}J^{c}\,,\qquad \ \ \ \left[
J_{a},P_{b}\right] =\epsilon _{abc}P^{c}\,,  \notag \\
\left[ P_{a},P_{b}\right] &=&\epsilon _{abc}Z^{c}\,,\qquad \ \ \ \left[
J_{a},Z_{b}\right] =\epsilon _{abc}Z^{c}\,, \notag\\
\left[ Z_{a},Z_{b}\right] &=&\frac{1}{\ell ^{2}}\epsilon
_{abc}Z^{c}\,,\qquad \left[ P_{a},Z_{b}\right] =\frac{1}{\ell ^{2}}\epsilon
_{abc}P^{c}\,,  \label{sadsL1}
\end{eqnarray}
\begin{eqnarray}
\left[ J_{a},Q_{\alpha }\right] &=&\frac{1}{2}\,\left( \Gamma _{a}\right) _{%
\text{ }\alpha }^{\beta }Q_{\beta }\,,\qquad \ \,\left[ J_{a},\Sigma
_{\alpha }\right] =\frac{1}{2}\,\left( \Gamma _{a}\right) _{\text{ }\alpha
}^{\beta }\Sigma _{\beta }\,,  \notag \\
\left[ P_{a},Q_{\alpha }\right] &=&\frac{1}{2}\,\,\left( \Gamma _{a}\right)
_{\text{ }\alpha }^{\beta }\Sigma _{\beta }\,,\qquad \ \,\left[ P_{a},\Sigma
_{\alpha }\right] =\frac{1}{2\ell ^{2}}\,\left( \Gamma _{a}\right) _{\text{ }%
\alpha }^{\beta }Q_{\beta }\,,  \notag \\
\left[ Z_{a},Q_{\alpha }\right] &=&\frac{1}{2\ell ^{2}}\,\left( \Gamma
_{a}\right) _{\text{ }\alpha }^{\beta }Q_{\beta }\,,\qquad \left[
Z_{a},\Sigma _{\alpha }\right] =\frac{1}{2\ell ^{2}}\,\left( \Gamma
_{a}\right) _{\text{ }\alpha }^{\beta }\Sigma _{\beta }\,, \notag\\
\left\{ Q_{\alpha },Q_{\beta }\right\} &=&-\frac{1}{2}\,\left( C\Gamma
^{a}\right) _{\alpha \beta }P_{a}\,,\text{ \ \ }\left\{ Q_{\alpha },\Sigma
_{\beta }\right\} =-\frac{1}{2}\,\left( C\Gamma ^{a}\right) _{\alpha \beta
}Z_{a}\,,\,  \notag \\
\left\{ \Sigma _{\alpha },\Sigma _{\beta }\right\} &=&-\frac{1}{2\ell ^{2}}%
\,\,\left( C\Gamma ^{a}\right) _{\alpha \beta }P_{a}\,,\label{sadsL2}
\end{eqnarray}
where $a,b,\dots =0,1,2$ are Lorentz indices raised and lowered with the (off-diagonal) Minkowski metric $\eta _{ab}$ and $\epsilon _{abc}$ is the Levi-Civita tensor. The gamma matrices in three dimensions are denoted by $\Gamma_a$ and $C$ is the charge conjugation matrix, satisfying $C^{T}=-C$ and $C\Gamma^{a}=(C\Gamma^{a})^{T}$. This supersymmetric extension of the AdS-Lorentz symmetry, as well as its flat limit $\ell \rightarrow \infty $, is characterized by the introduction of an additional Majorana spinor charge $\Sigma$. Let us note that the presence of a second abelian spinorial generator has already been studied in the context of $D=11$ supergravity \cite{DAuria:1982uck} and superstring theory \cite{Green:1989nn}.

We can see that the aforesaid limit $\ell \rightarrow \infty $ leads to
the minimal Maxwell superalgebra, denoted in \cite{Concha:2018jxx} as
$s\mathcal{M}$, and whose underlying symmetry corresponds to a supersymmetric extension of the Maxwell algebra \cite{Schrader:1972zd, Bacry:1970ye, Gomis:2017cmt}, which in turn corresponds to a deformation and extension of the Poincaré symmetry. As we will see later, the flat limit is also well-defined at the level of the asymptotic symmetries of the supergravity theories based on the Maxwell and AdS-Lorentz superalgebras.

There exist an alternative basis in which the minimal AdS-Lorentz superalgebra
can be written as the direct sum of three copies of the Lorentz algebra, two of which are augmented by supersymmetry \cite{Caroca:2019dds} . This structure is revealed after the following redefinitions,
\begin{align}
    Z_a & =\frac{1}{\ell^{2}}\left(J_{a}^{+}+J_{a}^{-}\right)\,, &
 P_a &=\frac{1}{\ell}\left(J_{a}^{+}-J_{a}^{-}\right)\,,&
J_{a}&=\hat{J}_{a}+J_{a}^{+}+J_{a}^{-}\,,
\end{align}
along with
\begin{align}
    Q_{\alpha}& =\frac{1}{\ell^{1/2}}\left(\Tilde{Q}_{\alpha}^{+}-i\Tilde{Q}_{\alpha}^{-}\right)\,, &
    \Sigma_{\alpha}& =\frac{1}{\ell^{3/2}}\left(\Tilde{Q}_{\alpha}^{+}+i\Tilde{Q}_{\alpha}^{-}\right)\,.
\end{align}
Then, the minimal AdS-Lorentz superalgebra can be written as the direct sum  $\mathfrak{osp}(2,1)\oplus\mathfrak{osp}(2,1)\oplus \mathfrak{sp}(2)$, where the $\mathfrak{osp}(2,1)$ algebras are spanned by $(J_{a}^{+},\Tilde{Q}_{\alpha}^{+})$, $(J_{a}^{-},\Tilde{Q}_{\alpha}^{- })$  and the Lorentz algebra is spanned by $\hat{J}_{a}$.

As it was shown in \cite{Concha:2018jxx}, the flat limit can also be performed at the level of the CS action, supersymmetry transformations and field equations.
A CS action for the minimal AdS-Lorentz superalgebra can be written by considering the connection one-form 
\begin{equation}
A=\omega ^{a}J_{a}+e^{a}P_{a}+\sigma ^{a}Z_{a}+\bar{\psi}Q+\bar{\xi}\Sigma
\,,  \label{1fP}
\end{equation}
where $\omega^{a}$ is the spin connection one-form, $e^{a}$ corresponds to
the dreibein one-form, $\sigma ^{a}$ is the gauge field one-form along $Z_{a}$,
while $\psi $ and\ $\xi $ are fermionic fields.
The corresponding curvature two-form is given by
\begin{equation}
F=R^{a}J_{a}+\hat{\mathcal{T}}^{a}P_{a}+\hat{\mathcal{F}}^{a}Z_{a}+\nabla \bar{\psi}%
Q+\nabla \bar{\xi}\Sigma \,,
\end{equation}%
where%
\begin{eqnarray}
R^{a} &=&d\omega ^{a}+\frac{1}{2}\epsilon ^{abc}\omega _{b}\omega _{c}\,,
\notag \\
\hat{\mathcal{T}}^{a} &=&de^{a}+\epsilon ^{abc}\omega _{b}e_{c}+\frac{1}{\ell ^{2}}%
\epsilon ^{abc}\sigma _{b}e_{c}+\frac{1}{4}i\bar{\psi}\Gamma ^{a}\psi +\frac{1%
}{4\ell ^{2}}i\bar{\xi}\Gamma ^{a}\xi \,, \notag \\
\hat{\mathcal{F}}^{a} &=&d\sigma ^{a}+\epsilon ^{abc}\omega _{b}\sigma _{c}+\frac{1%
}{2\ell ^{2}}\epsilon ^{abc}\sigma _{b}\sigma _{c}+\frac{1}{2}\epsilon
^{abc}e_{b}e_{c}+\frac{1}{2}i\bar{\psi}\Gamma ^{a}\xi \,,  \label{bosc2}
\end{eqnarray}%
and the fermionic components are given by
\begin{eqnarray}
\nabla \psi &=&d\psi +\frac{1}{2}\,\omega ^{a}\Gamma _{a}\psi +\frac{1}{%
2\ell ^{2}}\,\sigma ^{a}\Gamma _{a}\psi +\frac{1}{2\ell ^{2}}\,e^{a}\Gamma
_{a}\xi \,,  \notag \\
\nabla \xi &=&d\xi +\frac{1}{2}\,\omega ^{a}\Gamma _{a}\xi \,+\frac{1}{2\ell
^{2}}\,\sigma ^{a}\Gamma _{a}\xi +\frac{1}{2}\,e^{a}\Gamma _{a}\psi \,. \label{ferc2}
\end{eqnarray}%
Naturally, the curvatures reduce to the minimal Maxwell ones \cite{Concha:2018jxx} when the limit $\ell \rightarrow \infty $ is performed.

The non-vanishing components of the invariant tensor for the minimal AdS-Lorentz superalgebra are given by those for the Maxwell superalgebra, namely
\begin{align}
\left\langle J_{a}J_{b}\right\rangle &=\alpha _{0}\eta _{ab}\,,  &  \left\langle J_{a}P_{b}\right\rangle &=\alpha _{1}\eta _{ab}\,, & \left\langle P_{a}P_{b}\right\rangle &=\alpha _{2}\eta _{ab}\,,  \notag \\
\left\langle J_{a}Z_{b}\right\rangle &=\alpha _{2}\eta _{ab}\,, & 
\left\langle Q_{\alpha }Q_{\beta }\right\rangle &=\alpha _{1}C_{\alpha \beta
}\,, & \left\langle Q_{\alpha }\Sigma _{\beta }\right\rangle &=\alpha _{2}C_{\alpha
\beta}\,,\label{ITMax}
\end{align}
along with
\begin{align}
\left\langle P_{a}Z_{b}\right\rangle & =\frac{\alpha _{1}}{\ell ^{2}}\eta
_{ab}\,,  \notag \\
\left\langle Z_{a}Z_{b}\right\rangle & =\frac{\alpha _{2}}{\ell ^{2}}\eta
_{ab}\,,  \label{AdSLorInv} \\
\left\langle \Sigma _{\alpha }\Sigma _{\beta }\right\rangle & =\frac{\alpha
_{1}}{\ell ^{2}}C_{\alpha \beta }\,.  \notag
\end{align}
%Thus, the most general quadratic Casimir of the minimal AdS-Lorentz
%superalgebra is
%\begin{equation}
%C=\alpha_{0}J^{a}J_{a}+\alpha_{1}\left(P^{a}J_{a}+\frac{1}{\ell^{2}}%
%P^{a}Z_{a}+\bar{Q}Q+\frac{1}{\ell^{2}}\bar{\Sigma}\Sigma\right)+\alpha_2%
%\left(P^{a}P_{a}+J^{a}Z_{a}+\frac{1}{\ell^{2}}Z^{a}Z_{a}+\bar{Q}%
%\Sigma\right)\,.
%\end{equation}
We can see the flat limit $\ell \rightarrow \infty$ applied to the non-vanishing components  of the invariant tensor for the AdS-Lorentz superalgebra leads us to the non-degenerate invariant tensor of the minimal Maxwell superalgebra. Then, using the connection one-form (\ref{1fP}) and
the invariant tensors (\ref{ITMax}) and (\ref{AdSLorInv}) in the CS
action 
\begin{equation}
I[A]=\frac{k}{4\pi }\int_{\mathcal{M}}\left\langle AdA+\frac{2}{3}
A^{3}\right\rangle \,,  \label{CSaction}
\end{equation}
defined on a three-dimensional manifold $\mathcal{M}$, and where $k=\frac{1}{4G}$ is the level of the theory related to the gravitational constant $G$, we get \cite{Concha:2018jxx}
\begin{align}
I_{sAdS-\mathcal{L}}\,=\,& \frac{k}{4\pi }\int \alpha _{0}\left( \,\omega
^{a}d\omega _{a}+\frac{1}{3}\,\epsilon _{abc}\omega ^{a}\omega ^{b}\omega
^{c}\right)  \notag \\
& +\alpha _{1}\left( 2e^{a}R_{a}+\frac{2}{\ell ^{2}}\,e^{a}F_{a}+\frac{1}{%
3\ell ^{2}}\,\epsilon _{abc}e^{a}e^{b}e^{c}-i\bar{\psi}\nabla \psi -\frac{1}{%
\ell ^{2}}\,i\bar{\xi}\nabla \xi \right)  \notag \\
& +\alpha _{2}\left( 2R^{a}\sigma _{a}+\frac{2}{\ell ^{2}}\,F^{a}\sigma
_{a}+e^{a}T_{a}+\frac{1}{\ell ^{2}}\,\epsilon _{abc}e^{a}\sigma ^{b}e^{c}-%
i\bar{\psi}\nabla \xi -i\bar{\xi}\nabla \psi \right) \,,  \label{AdSLorAct}
\end{align}%
where $T^{a}=de^{a}+\epsilon ^{abc}\omega _{b}e_{c}$ is the usual torsion two-form and
\begin{equation}
F^{a}=d\sigma ^{a}+\epsilon ^{abc}\omega _{b}\sigma _{c}+\frac{1}{2\ell ^{2}}%
\epsilon ^{abc}\sigma _{b}\sigma _{c}\,.
\end{equation}
From the previous action we can see that the term proportional to $\alpha _{0}$ contains the gravitational CS 
Lagrangian \cite{Witten:1988hc,Zanelli:2005sa}. The term along $\alpha _{1}$ contains the Einstein-Hilbert term with negative cosmological constant and the Rarita-Schwinger term, while $\alpha _{2}$ yields a term involving the field $\sigma ^{a}$, a torsional term and supersymmetric terms involving the second spinorial gauge field. Naturally, in absence of supersymmetry, the CS action corresponds to the AdS-Lorentz CS gravity.
As we have mentioned before, the minimal AdS-Lorentz supergravity
action (\ref{AdSLorAct}) leads to the minimal Maxwell
supergravity in the vanishing cosmological constant limit $\ell \rightarrow \infty $. Although the field content in both theories is the same, the CS action constructed from the AdS-Lorentz superalgebra is richer in the sense that it allows to introduce a negative cosmological constant.
The field equations reduce to the vanishing of the curvature two-forms \eqref{bosc2} and \eqref{ferc2} provided $\alpha _{2}\neq 0$. This condition on the $\alpha _{2}$ constant is required, as in the bosonic case, in order to have a non-degenerate invariant tensor.
%\begin{eqnarray}
%\delta \omega ^{a} &:&\text{ \ \ \ }\alpha _{0}R_{a}+\alpha _{1}\mathcal{T}%
%_{a}+\alpha _{2}\mathcal{F}_{a}\,=0\,,  \notag \\
%\delta e^{a} &:&\text{ \ \ \ }\alpha _{1}\left( R_{a}+\frac{1}{\ell ^{2}}\,%
%\mathcal{F}_{a}\right) +\alpha _{2}\mathcal{T}_{a}=0\,,  \notag \\
%\delta \sigma ^{a} &:&\text{ \ \ }\frac{\alpha _{1}}{\ell ^{2}}\,\mathcal{T}%
%_{a}+\alpha _{2}\left( R_{a}+\frac{1}{\ell ^{2}}\,\mathcal{F}_{a}\right)
%\,=0\,,  \label{adsleq} \\
%\delta \bar{\psi} &:&\qquad \alpha _{1}\nabla \psi +\alpha _{2}\nabla \xi
%=0\,,  \notag \\
%\delta \bar{\xi} &:&\qquad \alpha _{2}\nabla \psi +\frac{\alpha _{1}}{\ell
%^{2}}\nabla \xi =0\,.  \notag
%\end{eqnarray}%
The CS action (\ref{AdSLorAct}) is invariant by construction under the gauge
transformation $\delta A=d\Lambda +\left[ A,\Lambda \right] $. In
particular, the supersymmetry transformation laws read%
\begin{eqnarray}
\delta \omega ^{a} &=&0\,,  \notag \\
\delta e^{a} &=&\frac{1}{2}\,i\bar{\epsilon}\Gamma ^{a}\psi \,+\frac{1}{2\ell
^{2}}i\,\bar{\varrho}\Gamma ^{a}\xi ,  \notag \\
\delta \sigma ^{a} &=&\frac{1}{2}\,i\bar{\epsilon}\Gamma ^{a}\xi +\frac{1}{2}%
\,i\bar{\varrho}\Gamma ^{a}\psi \,,  \label{Asusy} \\
\delta \psi &=&d\epsilon +\frac{1}{2}\,\omega ^{a}\Gamma _{a}\epsilon +\frac{%
1}{2\ell ^{2}}\sigma ^{a}\Gamma _{a}\epsilon +\frac{1}{2\ell ^{2}}%
e^{a}\Gamma _{a}\varrho \,,  \notag \\
\delta \xi &=&d\varrho +\frac{1}{2}\,\omega ^{a}\Gamma _{a}\varrho \,+\frac{1%
}{2}e^{a}\Gamma _{a}\epsilon +\frac{1}{2\ell ^{2}}\sigma ^{a}\Gamma
_{a}\varrho \,,  \notag
\end{eqnarray}%
with gauge parameter $\Lambda =\rho ^{a}J_{a}+\varepsilon ^{a}P_{a}+\gamma
^{a}Z_{a}+\bar{\epsilon}Q+\bar{\varrho}\Sigma $. Note that the supersymmetry
transformations (\ref{Asusy}) reduce to the Maxwell ones in the limit $\ell \rightarrow \infty $.

\ \

In what follows, we extend the results obtained in
\cite{Concha:2018zeb,Concha:2018jjj} to the minimal AdS-Lorentz supergravity theory and its flat limit, leading to the minimal Maxwell supergravity. As we will see, the corresponding asymptotic symmetry algebras of the supergravity theories will correspond to a supersymmetric extension of the semi-simple enlargement of the $\mathfrak{bms}
_{3}$ algebra in the former case, and to a deformed super-$\mathfrak{bms}
_{3}$ in the latter case. Both charge algebras have three central charges defined in terms of the coupling constants appearing in the CS action. We will see that both asymptotic symmetry algebras are related through a well-defined flat limit.

\section{Asymptotic symmetries}\label{sec3}
In this section, we compute the asymptotic symmetry algebra for the previously introduced minimal AdS-Lorentz CS supergravity. To start with, we provide the suitable fall-off
conditions for the gauge fields at infinity and the gauge transformations preserving our boundary
conditions. Then, the charge algebra is found using the Regge-Teitelboim method \cite{Regge:1974zd}.   

\subsection{Boundary conditions}

We propose the following behaviour of the gauge fields at the boundary 

\begin{equation}\label{AsMF}
A=h^{-1}dh+h^{-1}ah\,,
\end{equation}
where the radial dependence is entirely captured by the group element $h=$ e$^{-rP_{0}}$. The auxiliary gauge field has the form $a=a_{\phi}d{\phi} +a_udu$, with the angular component given by
\begin{eqnarray}
a_\phi &= &  J_{1}+\frac{1}{2}{\cal N} P_{0}+\frac{1}{2}\left({\cal  M}-\frac{{\cal F}}{\ell^2} \right)\, J_{0}+\frac{1}{2}\mathcal{F}\,Z_{0} +\frac{\psi}{2^{1/4}}\, Q_{+}+\frac{1}{2^{1/4}}\xi \,\Sigma_{+}\,,\label{aa}
\end{eqnarray}
where the functions ${\cal M}$, ${\cal N}$ and $\cal{F}$, and the Grassmann-valued spinor components $\psi$ and $\xi$ are assume to depend on all boundary coordinates $x^{i}=(u,\phi)$. 

The asymptotic symmetries correspond to the set of gauge transformations $\delta A = d\lambda + [A,\lambda]$
that preserve the asymptotic conditions (\ref{AsMF}) invariant. Thus, the angular component $a_\phi$ is left invariant for the Lie-algebra-valued parameter of the form

\begin{align}
\lambda[Y,f,h,\mathcal{E},\mathcal{S}] = &\, \left[\frac{1}{2}\left(\mathcal{M}Y+\mathcal{N} Y+\mathcal{N}\frac{h}{\ell^{2}}\right)+\frac{{i}}{2}\mathcal{E}\psi+\frac{{i}}{2\ell}\mathcal{S}\xi-f^{\prime\prime} \right] P_{0}+ f P_{1}- f^{\prime} P_{2} \nonumber \\
 &+\left[\frac{1}{2}\left(\mathcal{M}-\frac{\mathcal{F}}{\ell^{2}}\right)\,Y-Y^{\prime\prime} \right]\,J_{0}+Y\,J_{1}-Y^{\prime}\,J_{2} \nonumber \\
 &+\left[\frac{1}{2}\left(\mathcal{M}h+\mathcal{N}f+\mathcal{F}Y\right)+\frac{{i}}{2}\mathcal{S}\psi+\frac{{i}}{2}\mathcal{E}\xi-h^{\prime\prime}\right] Z_{0}+h Z_{1}-h^{\prime}Z_{2}\nonumber \\
 &+2^{-1/4}\left[\left(Y+\frac{h}{\ell^{2}}\right)\psi+\frac{f} {\ell^{2}}\xi-2\mathcal{E}^{\prime}\right]Q_{+} +2^{1/4}\mathcal{E} \,Q_{-} \nonumber \\
 &+2^{-1/4}\left[\left(Y+\frac{h}{\ell^{2}}\right)\xi+f\psi-2\mathcal{S}^{\prime}\right]\Sigma_{+}+2^{1/4}\mathcal{S}\Sigma_{-}\,,\label{lambda}
\end{align}
provided the dynamical fields transform in the following way
\begin{eqnarray}
\delta{\cal M} & = & {\cal M}^{\prime}Y+2{\cal M}Y^{\prime}-2Y{}^{\prime\prime\prime}+\frac{1}{\ell^{2}}\left({\cal M}^{\prime}h+2{\cal M}h^{\prime}+{\cal N}^{\prime}f+2{\cal N}f^{\prime}+i\mathcal{E}\Xi^{\prime}+3i\mathcal{E}^{\prime}\Xi+{i}\mathcal{S}\Psi^{\prime}+3i\mathcal{S}^{\prime}\Psi-2h{}^{\prime\prime\prime}\right)\,,\nonumber \\
\delta{\cal N} & = & {\cal N}^{\prime}Y+2{\cal N}Y^{\prime}+{\cal M}^{\prime}f+2{\cal M}f^{\prime}+{i}\mathcal{E}\Psi^{\prime}+3{i}\mathcal{E}^{\prime}\Psi-2f{}^{\prime\prime\prime}+\frac{1}{\ell^{2}}\left({\cal N}^{\prime}h+2{\cal N}h^{\prime}+i\mathcal{S}\Xi^{\prime}+3i\mathcal{S}^{\prime}\Xi\right)\,,\nonumber\\
\delta{\cal F} & = & {\cal F}^{\prime}Y+2{\cal F}Y^{\prime}+{\cal N}^{\prime}f+2{\cal N}f^{\prime}+{\cal M}^{\prime}h+2{\cal M}h^{\prime}+{i}\mathcal{E}\Xi^{\prime}+3{i}\mathcal{E}^{\prime}\Xi+{i}\mathcal{S}\Psi^{\prime}+3{i}\mathcal{S}^{\prime}\Psi-2h{}^{\prime\prime\prime}\,,\nonumber \\
\delta{ \Psi} & = &\Psi^{\prime}Y+\frac{3}{2}\Psi Y^{\prime}+\frac{1}{2}\mathcal{M}\mathcal{E}-2\mathcal{E}^{\prime\prime}+\frac{1}{\ell^2}\left(\Psi^{\prime}h+\frac{3}{2}\Psi h^{\prime}+\Xi^{\prime}f+\frac{3}{2}\Xi f^{\prime}+\frac{1}{2}\mathcal{N}\mathcal{S}\right)\,,\nonumber\\
\delta{\Xi} & = & \Xi^{\prime}Y+\frac{3}{2}\Xi Y^{\prime}+\Psi^{\prime}f+\frac{3}{2}\Psi f^{\prime}+\frac{1}{2}\mathcal{M}\mathcal{S}+\frac{1}{2}\mathcal{N}\mathcal{E}-2\mathcal{S}^{\prime\prime}+\frac{1}{\ell^2}\left(\Xi^{\prime}h+\frac{3}{2}\Xi h^{\prime}\right)\,.\label{transflawadsLorentz}
\end{eqnarray}

 In order to accommodate the relevant bosonic solutions of the theory, it is important to incorporate the Lagrange multipliers for every dynamical field in the asymptotic form of the gauge field. Therefore, the most general form of the temporal component is given by
\begin{equation}\label{AsMu}
a_u=\lambda[\mu_{\mathcal{N}}, \mu_{\mathcal{M}},\mu_{\mathcal{F}},\mu_{\Psi},\mu_{\Xi}]\,,
\end{equation}
where the lagrange multipliers $\mu_{\mathcal{N}}$, $\mu_{\mathcal{M}}$, $\mu_{\mathcal{F}}$, $\mu_{\Psi}$, $\mu_{\Xi}$ are arbitrary functions of $(u,\phi)$ and also assumed to be fixed at the boundary \cite{Henneaux:2013dra, Bunster:2014mua}. The preservation of the temporal component \eqref{AsMu} requires that the field equations hold asymptotically as well as they provide conditions for the temporal dependence of the gauge parameter. The explicit form of these conditions are given by
\begin{eqnarray}
\dot{\cal M} & = & {\cal M}^{\prime}\mu_{\mathcal{N}}+2{\cal M}\mu_{\mathcal{N}}^{\prime}-2\mu_{\mathcal{N}}^{\prime\prime\prime}+\frac{1}{\ell^{2}}\left({\cal M}^{\prime}\mu_{\mathcal{F}}+2{\cal M}\mu_{\mathcal{F}}^{\prime}+{\cal N}^{\prime}\mu_{\mathcal{M}}+2{\cal N}\mu_{\mathcal{M}}^{\prime}+i\mu_{\Psi}\Xi^{\prime}+3i\mu_{\Psi}^{\prime}\Xi \right.\,\nonumber \\
&&\left.+{i}\mu_{\Xi}\Psi^{\prime}+3i\mu_{\Xi}^{\prime}\Psi-2\mu_{\mathcal{F}}{}^{\prime\prime\prime}\right)\,, \nonumber \\
\dot{\cal N} & = & {\cal N}^{\prime}\mu_{\mathcal{N}}+2{\cal N}\mu_{\mathcal{N}}^{\prime}+{\cal M}^{\prime}\mu_{\mathcal{M}}+2{\cal M}\mu_{\mathcal{M}}^{\prime}+{i}\mu_{\Psi}\Psi^{\prime}+3{i}\mu_{\Psi}^{\prime}\Psi-2\mu_{\mathcal{M}}{}^{\prime\prime\prime}+\frac{1}{\ell^{2}}\left({\cal N}^{\prime}\mu_{\mathcal{F}}+2{\cal{N}}\mu_{\mathcal{F}}^{\prime}\right. \nonumber \\
&&\left.+i\mu_{\Xi}\Xi^{\prime}+3i\mu_{\Xi}^{\prime}\Xi\right)\,,\nonumber \\
\dot{\cal F} & = & {\cal F}^{\prime}\mu_{\mathcal{N}}+2{\cal F}\mu_{\mathcal{N}}^{\prime}+{\cal N}^{\prime}\mu_{\mathcal{M}}+2{\cal N}\mu_{\mathcal{M}}^{\prime}+{\cal M}^{\prime}\mu_{\mathcal{F}}+2{\cal M}\mu_{\mathcal{F}}^{\prime}+{i}\mu_{\psi}\Xi^{\prime}+3{i}\mu_{\psi}^{\prime}\Xi+{i}\mu_{\Xi}\Psi^{\prime}+3{i}\mu_{\Xi}^{\prime}\Psi-2\mu_{\mathcal{F}}^{\prime\prime\prime}\,,\nonumber \\
\dot{ \Psi} & = &\Psi^{\prime}\mu_{{\cal N}}+\frac{3}{2}\Psi \mu_{{\cal N}}^{\prime}+\frac{1}{2}\mathcal{M}\mu_{\Psi}-2\mu_{\Psi}^{\prime\prime}+\frac{1}{\ell^2}\left(\Psi^{\prime}\mu_{\mathcal{F}}+\frac{3}{2}\Psi \mu_{\mathcal{F}}^{\prime}+\Xi^{\prime}\mu_{\mathcal{M}}+\frac{3}{2}\Xi \mu_{\mathcal{M}}^{\prime}+\frac{1}{2}\mathcal{N}\mu_{\Xi}\right)\,,\nonumber\\
\dot{\Xi} & = & \Xi^{\prime}\mu_{\mathcal{N}}+\frac{3}{2}\Xi \mu_{\mathcal{N}}^{\prime}+\Psi^{\prime}\mu_{\mathcal{M}}+\frac{3}{2}\Psi \mu_{\mathcal{M}}^{\prime}+\frac{1}{2}\mathcal{M}\mu_{\Xi}+\frac{1}{2}\mathcal{N}\mu_{\psi}-2\mu_{\Xi}^{\prime\prime}+\frac{1}{\ell^2}\left(\Xi^{\prime}\mu_{\mathcal{F}}+\frac{3}{2}\Xi \mu_{\mathcal{F}}^{\prime}\right)\,.\label{transflawnew}
\end{eqnarray}
% \begin{eqnarray}
% \dot{\cal M} & = & {\cal M}^{\prime}\mu_{\mathcal{J}}+2{\cal M}\mu_{\mathcal{J}}^{\prime}-2\mu_{\mathcal{J}}^{\prime\prime\prime}\,,\nonumber \\
% \dot{\cal J} & = & {\cal J}^{\prime}\mu_{{\cal J}}+2{\cal J}\mu_{{\cal J}}^{\prime}+{\cal M}^{\prime}\mu_{{\cal M}}+2{\cal M}\mu_{{\cal M}}^{\prime}+i\mu_{\Psi}\Psi^{\prime}+3i\mu_{\Psi}^{\prime}\Psi-2\mu_{{\cal M}}^{\prime\prime\prime}\,,\nonumber\\
% \dot{\cal Z} & = & {\cal Z}^{\prime}\mu_{\mathcal{J}}+2{\cal Z}\mu_{\mathcal{J}}^{\prime}+{\cal J}^{\prime}\mu_{\mathcal{M}}+2{\cal J}\mu_{\mathcal{M}}^{\prime}+{\cal M}^{\prime}\mu_{\mathcal{Z}}+2{\cal M}\mu_{\mathcal{Z}}^{\prime}+\textcolor{red}{i}\mu_{\psi}\Xi^{\prime}+3\textcolor{red}{i}\mu_{\psi}^{\prime}\Xi+\textcolor{red}{i}\mu_{\Xi}\Psi^{\prime}+3\textcolor{red}{i}\mu_{\Xi}^{\prime}\Psi-2\mu_{\mathcal{Z}}^{\prime\prime\prime}\,,\nonumber \\
% \dot{ \Psi} & = &\Psi^{\prime}\mu_{{\cal J}}+\frac{3}{2}\Psi \mu_{{\cal J}}^{\prime}+\frac{1}{2}\mathcal{M}\mu_{\Psi}-2\mu_{\Psi}^{\prime\prime}\,,\nonumber\\
% \dot{\Xi} & = & \Xi^{\prime}\mu_{\mathcal{J}}+\frac{3}{2}\Xi \mu_{\mathcal{J}}^{\prime}+\Psi^{\prime}\mu_{\mathcal{M}}+\frac{3}{2}\Psi \mu_{\mathcal{M}}^{\prime}+\frac{1}{2}\mathcal{M}\mu_{\Xi}+\frac{1}{2}\mathcal{J}\mu_{\psi}-2\mu_{\Xi}^{\prime\prime}\,.\label{transflaw}
% \end{eqnarray}
for the asymptotic field, and for the parameters we have
\begin{align}\label{flatcond}
\dot{Y} &= Y^{\prime}\mu_{\mathcal{J}} - \mu_{\mathcal{J}}^{\prime} Y \nonumber \\
\dot{f} &= Y^{\prime}\mu_{\mathcal{M}} - \mu_{\mathcal{M}}^{\prime} Y +f^{\prime}\mu_{\mathcal{J}} - \mu_{\mathcal{J}}^{\prime} f -i \nonumber \mu_{\Psi}\mathcal{E} + \frac{1}{\ell^2}\left( f^{\prime} \mu_{Z} -\mu_{Z}^{\prime} f + h^{\prime} \mu_{M} -\mu_{M}^{\prime} h -\mu_{\Xi} \mathcal{S} \right) \nonumber\\
\dot{h} &= Y^{\prime}\mu_{\mathcal{Z}} - \mu_{\mathcal{Z}}^{\prime} Y + f^{\prime}\mu_{\mathcal{M}} - \mu_{\mathcal{M}}^{\prime} f+ h^{\prime}\mu_{\mathcal{J}} - \mu_{\mathcal{J}}^{\prime} h -\mu_{\Xi} \mathcal{E} - \mu_{\psi} \mathcal{S}+\frac{1}{\ell^2}\left(h^{\prime}\mu_{Z} - \mu_{Z}^{\prime}h \right)\nonumber \\
\dot{\mathcal{E}} &= \mu_{\mathcal{J}} \mathcal{E}^{\prime}- \frac{1}{2} \mathcal{E} \mu_{\mathcal{J}}^{\prime}- \mu_{\Psi}^{\prime}Y+\frac{1}{2} \mu_{\Psi} Y^{\prime} +\frac{1}{\ell^2} \left( \mu_{M} \mathcal{S}^{\prime}-\frac{1}{2} \mathcal{S} \mu_{M}^{\prime}+\mu_{Z} \mathcal{E}^{\prime}-\frac{1}{2} \mathcal{E} \mu_{Z}^{\prime} -f \mu_{\Sigma}^{\prime} + \frac{1}{2} \mu_{\Sigma} f^{\prime} - h \mu_{\psi}^{\prime} +  \frac{1}{2} \mu_{\psi} h^{\prime}\right)  \nonumber \\
\dot{\mathcal{S}} &=\mu_{\mathcal{J}} \mathcal{S}^{\prime}- \frac{1}{2} \mathcal{S} \mu_{\mathcal{J}}^{\prime}+\mu_{\mathcal{M}} \mathcal{E}^{\prime}- \frac{1}{2} \mathcal{E} \mu_{\mathcal{M}}^{\prime}- \mu_{\Xi}^{\prime}Y+\frac{1}{2} \mu_{\Xi} Y^{\prime}- \mu_{\psi}^{\prime}T+\frac{1}{2} \mu_{\psi} T^{\prime} +\frac{1}{\ell^2} \left( \mu_{Z} \mathcal{S}^{\prime} -\frac{1}{2} \mathcal{S} \mu_{Z}^{\prime}+\mu_{\Xi} h^{\prime} -\frac{1}{2} h \mu_{\Xi}^{\prime}\right) \, .
\end{align}
 
The above structure contains the information of the asymptotic structure of the minimal AdS-Lorentz CS supergravity and their corresponding algebra. Indeed, the charge algebra of the theory can be computed following the Regge-Teitelboim approach \cite{Regge:1974zd}. In what follows we will consider this construction.

\subsection{Charge algebra}
Let us now compute the charge algebra of the supergravity theory previously introduced. As discussed in \cite{Banados:1994tn}, the algebra is spanned by the conserved charges
$Q[\Lambda]$. Furthermore, the charge algebra in representation of Poisson brackets can be obtained using the Regge-Teitelboim method directly from the transformation law
\begin{equation}\label{rt}
\delta_{\Lambda_{2}}Q[\Lambda_{1}]=\left\{ Q[\Lambda_{1}],Q[\Lambda_{2}]\right\} .
\end{equation}
On the other hand, the variation of the charge in CS theory is given
by 
\begin{equation}
\delta Q[\Lambda]=\frac{k}{2\pi}\int\limits _{\partial\Sigma}\left\langle \Lambda\delta A\right\rangle \,.
\end{equation}
After applying the gauge transformation \eqref{AsMF} which introduces the asymptotic field (\ref{aa}) we get
\begin{equation}
\delta Q[\lambda]=\frac{k}{2\pi}\int d\phi\left\langle \lambda\delta a_{\phi}\right\rangle \,. \label{QQ}
\end{equation}
Considering the invariant tensors \eqref{ITMax} and \eqref{AdSLorInv}, and the gauge field $a$ defined in \eqref{aa} in the previous expression, we get 
\begin{align}
\delta Q[Y,f,h,\mathcal{E},\mathcal{S}] =\frac{k}{4\pi}\int & d\phi\left[Y\left(\alpha_{0}{\delta \cal M}+\left(\alpha_{2}-\frac{\alpha_0}{\ell^{2}}\right)\delta{\cal Z}+\alpha_{1}\delta{\cal N}\right)+f\left(\alpha_{2}{\delta\cal N}+\alpha_{1}\delta{\cal M}\right)\right. \nonumber \\
& \left.+h\left(\alpha_{2}{\delta\cal M}+\frac{\alpha_1\delta\mathcal{N}}{\ell^2}\right)-2{i}\mathcal{E}\left(\alpha_{1}\delta\Psi+\alpha_{2}\delta\Xi\right)-2{i}\mathcal{S}\left(\alpha_{2}\delta\Psi+\frac{\alpha_1}{\ell^{2}}\delta\Xi\right)\right]\,.
\end{align}
We assume that the functions $Y$, $f$, $h$, $\mathcal{E}$ and $\mathcal{S}$ do not depend on the fields, in which case
it is trivial to integrate the variation out, finding
\begin{align}
Q[Y,f,h,\mathcal{E},\mathcal{S}] =\frac{k}{4\pi}\int & d\phi\left[Y\left(\alpha_{0}{ \cal M}+\left(\alpha_{2}-\frac{\alpha_0}{\ell^{2}}\right){\cal Z}+\alpha_{1}{\cal N}\right)+f\left(\alpha_{2}{\cal N}+\alpha_{1}{\cal M}\right)\right. \nonumber \\
& \left.+h\left(\alpha_{2}{\cal M}+\frac{\alpha_1\mathcal{N}}{\ell^2}\right)-2{i}\mathcal{E}\left(\alpha_{1}\Psi+\alpha_{2}\Xi\right)-2{i}\mathcal{S}\left(\alpha_{2}\Psi+\frac{\alpha_1}{\ell^{2}}\Xi\right)\right]\,.\label{QQ2}
\end{align}
Now we define the asymptotic charges which correspond to the independent terms in (\ref{QQ2}),
\begin{eqnarray}
j[Y] & = &\frac{k}{4\pi}\int d\phi\,Y\left(\alpha_{0}{\cal M}+\left(\alpha_{2}-\frac{\alpha_0}{\ell^{2}}\right){\cal Z}+\alpha_{1}{\cal N}\right)\,,\nonumber \\
p[f] & = & \frac{k}{4\pi}\int d\phi\,f\left(\alpha_{2}{\cal N}+\alpha_{1}{\cal M}\right)\,, \nonumber\\
z[h] & = & \frac{k}{4\pi}\int d\phi\,h\left(\alpha_{2}{\cal M}+\frac{\alpha_1\mathcal{N}}{\ell^2}\right)\,,\nonumber \\
g[\mathcal{E}] & = & \frac{k}{4\pi}\int d\phi{i}\,\mathcal{E}\left(\alpha_{1}\Psi+\alpha_{2}\Xi\right)\,, \nonumber \\
h[S] & = & \frac{k}{4\pi}\int d\phi{i}\,\mathcal{S}\left(\alpha_{2}\Psi+\frac{\alpha_1}{\ell^{2}}\Xi\right)\,.
\end{eqnarray}
Then, the Poisson brackets of these independent charges can be evaluated using (\ref{rt}) and (\ref{transflawadsLorentz}). Expanding in Fourier modes,
%\begin{eqnarray}
%\left\{ j[Y_{1}],j[Y_{2}]\right\}  & = & j\left[[Y_{1},Y_{2}]\right]+\frac{k\alpha_{0}}{2\pi}\int d\phi\,Y_{1}Y_{2}^{\prime\prime\prime}\,,\nonumber \\
%\left\{ j[Y],p[T]\right\}  & = & p\left[[Y,T]\right]+\frac{k}{2\pi}\int d\phi\,YT^{\prime\prime\prime}\,,\nonumber \\
%\left\{ j[Y],z[R]\right\}  & = & z\left[[Y,R]\right]+\frac{k\alpha_{2}}{2\pi}\int d\phi\,YR^{\prime\prime\prime}\,, \nonumber\\
%\left\{ p[T_{1}],p[T_{2}]\right\}  & = & z\left[[T_{1},T_{2}]\right]+\frac{k\alpha_{2}}{2\pi}\int d\phi\,T_{1}T_{2}^{\prime\prime\prime}\,,\nonumber \\
%\left\{ j[Y],q[\mathcal{E}]\right\}  & = & \,,\nonumber \\
%\left\{ p[T],q[\mathcal{E}]\right\} & = & \,,\nonumber \\
%\left\{ j[Y],g[\mathcal{S}]\right\}  & = & \,,\nonumber \\
%\left\{ q[\mathcal{E}_{1}],q[\mathcal{E}_{2}]\right\}  & = & \,,\nonumber \\
%\left\{ q[\mathcal{E}],g[\mathcal{S}\right\}  & = & \,,
%\end{eqnarray}
%where here $[x,y]=xy^{\prime}-yx^{\prime}$ stands for the Lie bracket of the vector
%field components $x(\phi)$ and $y\left(\phi\right)$ on $\partial\Sigma$.
%The result describes the minimal supersymmetric extension of the deformed $\mathfrak{bms}_{3}$ algebra introduced in \cite{Concha:2018zeb}. As it is
%expected, it corresponds to an infinite-dimensional lift of the minimal Maxwell
%superalgebra, with three central charges. 
\begin{equation}
{\cal J}_{m}=j[e^{im\text{\ensuremath{\phi}}}]\text{\,,\qquad}{\cal P}_{m}=p[e^{im\text{\ensuremath{\phi}}}]\text{\,,\qquad}{\cal Z}_{m}=z[e^{im\text{\ensuremath{\phi}}}]\text{\,,\qquad}{\cal G}_{m}=g[e^{im\text{\ensuremath{\phi}}}]\text{\,,\qquad}{\cal H}_{m}=h[e^{im\text{\ensuremath{\phi}}}]\,,
\end{equation}
the Poisson brackets read
\begin{align}
 i\left\{ \mathcal{J}_{m},\mathcal{J}_{n}\right\} & =  \left(m-n\right)\mathcal{J}_{m+n}+\dfrac{c_{1}}{12}\,m^{3}\delta_{m+n,0}\,,\nonumber\\
i\left\{ \mathcal{J}_{m},\mathcal{P}_{n}\right\}  & =  \left(m-n\right)\mathcal{P}_{m+n}+\dfrac{c_{2}}{12}\,m^{3}\delta_{m+n,0}\,,\nonumber\\
i\left\{ \mathcal{P}_{m},\mathcal{P}_{n}\right\}  & =  \left(m-n\right)\mathcal{Z}_{m+n}+\dfrac{c_{3}}{12}\,m^{3}\delta_{m+n,0}\,,\nonumber\\
i\left\{ \mathcal{J}_{m},\mathcal{Z}_{n}\right\}  & =  \left(m-n\right)\mathcal{Z}_{m+n}+\dfrac{c_{3}}{12}\,m^{3}\delta_{m+n,0}\,,\nonumber\\
i\left\{ \mathcal{P}_{m},\mathcal{Z}_{n}\right\} & =  \frac{1}{\ell^{2}}\left(m-n\right)\mathcal{P}_{m+n}+\dfrac{c_{2}}{12\ell^{2}}\,m^{3}\delta_{m+n,0}\,,\nonumber\\
i\left\{ \mathcal{Z}_{m},\mathcal{Z}_{n}\right\} & =  \frac{1}{\ell^{2}}\left(m-n\right)\mathcal{Z}_{m+n}+\dfrac{c_{3}}{12\ell^{2}}\,m^{3}\delta_{m+n,0}\,,\label{asymAdSL}
\end{align}
along with
\begin{align}
 i\left\{ \mathcal{J}_{m},\mathcal{G}_{n}\right\} &=\left( \frac{m}{2}-n\right)
\mathcal{G}_{m+n}\,, &  i\left\{ \mathcal{P}_{m},\mathcal{G}_{n}\right\}
&=\left( \frac{m}{2}-n\right) \mathcal{H}_{m+n}\,,  \notag \\
 i\left\{ \mathcal{J}_{m},\mathcal{H}_{n}\right\} &=\left( \frac{m}{2}-n\right)
\mathcal{H}_{m+n}\,, &  i\left\{ \mathcal{P}_{m},\mathcal{H}_{n}\right\}& =
\dfrac{1}{\ell^{2}}\left( \frac{m}{2}-n\right) \mathcal{G}_{m+n}\,,  \notag
\\
 i\left\{ \mathcal{Z}_{m},\mathcal{G}_{n}\right\} &=\dfrac{1}{\ell^{2}}\left(
\frac{m}{2}-n\right) \mathcal{G}_{m+n}\,,&  i\left\{ \mathcal{Z}_{m},\mathcal{H}_{n}\right\}&=\dfrac{1}{\ell ^{2}}\left( \frac{m}{2}-n\right)
\mathcal{H}_{m+n}\,,  \notag \\
i\left\{ \mathcal{G}_{m},\mathcal{G}_{n}\right\}  & = \mathcal{P}_{m+n}+\dfrac{c_{2}}{6}\,m^{2}\delta_{m+n,0}\,,\nonumber\\
i\left\{ \mathcal{G}_{m},\mathcal{H}_{n}\right\}  & = \mathcal{Z}_{m+n}+\dfrac{c_{3}}{6}\,m^{2}\delta_{m+n,0}\,, \notag \\
i\left\{ \mathcal{H}_{m},\mathcal{H}_{n}\right\}  & = \frac{1}{\ell^2}\mathcal{P}_{m+n}+\dfrac{c_{2}}{6\ell^{2}}\,m^{2}\delta_{m+n,0}\,,\label{asymAdSL2}
\end{align}
where we have used the integral representation of the Kronecker delta $\delta_{m,n}=\frac{1}{2\pi}\int d\phi\,e^{i(m-n)\text{\ensuremath{\phi}}}$. The previous algebra defines a supersymmetric extension of the semi-simple enlargement of the $\mathfrak{bms}_{3}$ algebra \cite{Concha:2018jjj}, and it was first introduced in \cite{Caroca:2019dds} as an expansion of the super Virasoro algebra. As it is expected, the asymptotic symmetry algebra corresponds to an infinite-dimensional lift of the minimal AdS-Lorentz superalgebra, with three central charges which are related to the CS level $k$ and to the three coupling constants appearing in the supergravity action through
\begin{equation}
    c_{i}=12 k\alpha_{i-1}\,, \qquad i=1,2,3
\end{equation}
Indeed, the minimal AdS-Lorentz superalgebra is a finite subalgebra of (\ref{asymAdSL})-(\ref{asymAdSL2})
with generators $\left\{ {\cal J_{{\rm -1}}{\rm ,}J_{{\rm 0}}{\rm ,{\cal J}}_{{\rm 1}}{\rm ,{\cal P}_{-1},{\cal P}_{0}}{\rm ,{\cal P}_{1}}{\rm ,{\cal Z}_{-1}}{\rm ,}{\cal Z}_{{\rm 0}}{\rm ,}{\cal Z}}_{1}{\rm ,}\cal Q_{{\rm 1/2}}{\rm ,}\cal Q_{{\rm -1/2}}{\rm ,}\cal G_{{\rm 1/2}}{\rm ,}\cal G_{{\rm -1/2}}\right\} $. This can be explicitly seen by identifying the modes in (\ref{asymAdSL})-(\ref{asymAdSL2}) with the generators in \eqref{sadsL1}-\eqref{sadsL2} as follows
\begin{align}
    \mathcal{J}_{-1}&=-\sqrt{2}J_{0}\,, & \mathcal{J}_{1}&=\sqrt{2}J_{1}\,, &   \mathcal{J}_{0}&=J_{2}\,, \nonumber
    \\
    \mathcal{P}_{-1}&=-\sqrt{2}P_{0}\,, & \mathcal{P}_{1}&=\sqrt{2}P_{1}\,, &   \mathcal{P}_{0}&=P_{2}\,, \nonumber \\
\mathcal{Z}_{-1}&=-\sqrt{2}Z_{0}\,, & \mathcal{Z}_{1}&=\sqrt{2}Z_{1}\,, &   \mathcal{Z}_{0}&=Z_{2}\,, \nonumber\\
  \mathcal{G}_{-1/2}&=\sqrt{2}Q_{+}\,, & \mathcal{G}_{1/2}&=\sqrt{2}Q_{-}\,, \nonumber\\  \mathcal{H}_{-1/2}&=\sqrt{2}\Sigma_{+}\,, &  \mathcal{H}_{1/2}&=\sqrt{2}\Sigma_{-}\,.
\end{align}
As it was mentioned in \cite{Caroca:2019dds}, the algebra (\ref{asymAdSL})-(\ref{asymAdSL2}) is isomorphic to the direct product of three copies of the Virasoro algebra, two of which are augmented by supersymmetry. Indeed, after the following redefinitions:

\begin{align}
  \mathcal{L}_{m}^{+}  &=\frac{1}{2}\left(\ell^{2}\mathcal{Z}_m +\ell \mathcal{P}_m\right)\,, & \mathcal{L}_{m}^{-}  &=\frac{1}{2}\left(\ell^{2}\mathcal{Z}_{-m} -\ell \mathcal{P}_{-m}\right)\,,&\hat{\mathcal{L}}_{m}=\mathcal{J}_{-m}-\ell^{2}\mathcal{Z}_{-m}\,, \notag \\
Q_{r}^{+}&=\frac{1}{2}\left(\ell^{1/2}\mathcal{G}_r +\ell^{3/2} \mathcal{H}_r\right)\,, & {Q}_{r}^{-}&=\frac{i}{2}\left(\ell^{1/2}\mathcal{G}_r -\ell^{3/2} \mathcal{H}_r\right)\,,\notag \\
  c^{\pm}&=\frac{1}{2}\left(\ell^{2}c_3\pm\ell c_2\right)\,, & \hat{c}&=\left(c_1-\ell^{2}c_3\right)\,.
\end{align}
the direct product $\mathfrak{svir}\oplus\mathfrak{svir}\oplus\mathfrak{vir}$ is revealed, 
\begin{equation}
\begin{array}{lcl}
i\left\{ \mathcal{L}_{m}^{\pm},\mathcal{L}_{n}^{\pm}\right\}  & = & \left(
m-n\right) \mathcal{L}_{m+n}^{\pm}+\dfrac{c^{\pm}}{12}\,m^3
\delta _{m+n,0}\,, \\[5pt]
i\left\{ \hat{\mathcal{L}}_{m},\hat{\mathcal{L}}_{n}\right\}  & = & \left(
m-n\right) \mathcal{\hat{L}}_{m+n}+\dfrac{\hat{c}}{12}\,m^3\delta _{m+n,0}\,, \\ [5pt]
i\left\{\mathcal{L}_{m}^{+},\mathcal{Q}_{r}^{+}\right\} & = & \left( \dfrac{m}{2}%
-r\right) \mathcal{Q}_{m+r}^{+}\,\,, \\[5pt]
i\left\{ \mathcal{L}_{m}^{-},\mathcal{Q}_{r}^{-}\right\} & = & \left( \dfrac{m
}{2}-r\right) \mathcal{Q}_{m+r}^{-}\,\,, \\[5pt]
i\left\{ \mathcal{Q}_{r}^{+},\mathcal{Q}_{s}^{+}\right\} & = & \mathcal{L}_{r+s}^{+}+
\dfrac{c^{+}}{6}\, r^{2} \delta _{r+s,0}\,, \\[5pt]
i\left\{ \mathcal{Q}_{r}^{-},\mathcal{Q}_{s}^{-}\right\} & = & \mathcal{L}^{-}
_{r+s}+\dfrac{c^{-}}{6}\,r^{2} \delta
_{r+s,0}\,.
\end{array}
\end{equation}

%\mathfrak{bms}

\section{Minimal deformed super-$\mathfrak{bms}_{3}$ algebra from a flat limit }
 As we have previously mentioned, Maxwell CS gravity can be obtained as a vanishing cosmological constant limit from the so-called AdS-Lorentz gravity in three-dimensions. Furthermore, considering the analysis of its corresponding asymptotic symmetries, it was shown in \cite{Concha:2018jjj} that the deformed $\mathfrak{bms}_{3}$ algebra, being the asymptotic symmetry of the Maxwell CS gravity, can be derived as a flat limit of a semi-simple enlargement of the $\mathfrak{bms}_{3}$ symmetry, which in turn corresponds to the asymptotic symmetry of the AdS-Lorentz CS gravity.

As we have discussed in the previous sections, in presence of supersymmetry, it was also shown that the minimal Maxwell supergravity appears as flat limit of a supersymmetric extension of the AdS-Lorentz algebra \cite{Concha:2018jxx}. In the same way, it is possible to show that a minimal deformed super-$\mathfrak{bms}_{3}$ algebra can be obtained from the enlarged super-$\mathfrak{bms}_{3}$ symmetry obtained above. Indeed, the vanishing cosmological constant limit $\ell\rightarrow\infty$ can be explicitly taken at every step in section \ref{sec3}, thus leading to the following asymptotic symmetry algebra 
\begin{align}
 i\left\{ \mathcal{J}_{m},\mathcal{J}_{n}\right\} & =  \left(m-n\right)\mathcal{J}_{m+n}+\dfrac{c_{1}}{12}\,m^{3}\delta_{m+n,0}\,,\nonumber\\
i\left\{ \mathcal{J}_{m},\mathcal{P}_{n}\right\}  & =  \left(m-n\right)\mathcal{P}_{m+n}+\dfrac{c_{2}}{12}\,m^{3}\delta_{m+n,0}\,,\nonumber\\
i\left\{ \mathcal{P}_{m},\mathcal{P}_{n}\right\}  & =  \left(m-n\right)\mathcal{Z}_{m+n}+\dfrac{c_{3}}{12}\,m^{3}\delta_{m+n,0}\,,\nonumber\\
i\left\{ \mathcal{J}_{m},\mathcal{Z}_{n}\right\}  & =  \left(m-n\right)\mathcal{Z}_{m+n}+\dfrac{c_{3}}{12}\,m^{3}\delta_{m+n,0}\,,\nonumber\\
i\left\{ \mathcal{J}_{m},\mathcal{G}_{n}\right\}  & =  \left(\frac{m}{2}-n\right)\mathcal{G}_{m+n}\,,\nonumber\\
i\left\{ \mathcal{P}_{m},\mathcal{G}_{n}\right\}  & = \left(\frac{m}{2}-n\right)\mathcal{H}_{m+n}\,,\nonumber\\
i\left\{ \mathcal{J}_{m},\mathcal{H}_{n}\right\}  & = \left(\frac{m}{2}-n\right)\mathcal{H}_{m+n}\,,\nonumber\\
i\left\{ \mathcal{G}_{m},\mathcal{G}_{n}\right\}  & = \mathcal{P}_{m+n}+\dfrac{c_{2}}{6}\,m^{2}\delta_{m+n,0}\,,\nonumber\\
i\left\{ \mathcal{G}_{m},\mathcal{H}_{n}\right\}  & = \mathcal{Z}_{m+n}+\dfrac{c_{3}}{6}\,m^{2}\delta_{m+n,0}\,, \label{asymminmax}
\end{align}
corresponding to a supersymmetric extension of the deformed
$\mathfrak{bms}_{3}$ algebra \cite{Caroca:2019dds,Concha:2018zeb}. It is worth highlighting that (\ref{asymminmax}) corresponds to the asymptotic symmetry algebra of the minimal Maxwell supergravity theory in three dimensions.

%%%%%%%%%%%%%%%%%%%%%%%%%%%%%%%%
\section{Discussion}\label{concl}

In this paper, we have studied the asymptotic symmetries of the three-dimensional Chern-Simons supergravity theories based on the supersymmetric extension of two different algebras connected by a flat limit, i.e. the AdS-Lorentz and the Maxwell algebras.  When imposing appropriate boundary conditions, we found that our results describe new asymptotic
structures of these supergravity theories. Similarly to the AdS-Lorentz and Maxwell superalgebras, the asymptotic generators
contain additional generators $\mathcal{Z}_{m}$ along with fermionic charges 
which modify the super-$\mathfrak{bms\mathrm{_{3}}}$ symmetry of the simplest $\mathcal{N}=1$ flat supergravity. In the AdS-Lorentz case, the asymptotic symmetry algebra is given by an enlarged super- $\mathfrak{bms}\mathrm{_3}$ symmetry. Although this algebra was already presented in \cite{Caroca:2019dds}, here we have shown that it is indeed the asymptotic symmetry algebra of the AdS-Lorentz supergravity theory in three space-time dimensions. The same statement is valid when the flat limit is considered. Indeed, the flat limit can be performed throughout all the steps followed in obtaining the asymptotic symmetry algebra for the AdS-Lorentz supergravity. In that case, the obtained charge algebra would have been the deformed super- $\mathfrak{bms}\mathrm{_3}$. Then,  our findings extend the results of \cite{Concha:2018jjj}, where it was explored the flat Maxwell limit at the level of the asymptotic boundary conditions.

Let us mention that the solutions of the Maxwell and AdS-Lorentz gravity theories were partially studied in \cite{Concha:2018zeb} and \cite{Concha:2018jjj}, respectively. Nonetheless, in order to consider an exhaustive analysis of the energy bounds and asymptotic Killing spinors it is necessary to fully understand these solutions and the thermodynamics. We will leave this discussion for a future work which is already in progress.

It would be worth it to extend our results to the case of $\mathcal{N}$-supersymmetries. One could expect that the asymptotic symmetry of the $\mathcal{N}$-extended Maxwell CS supergravity corresponds to a $\mathcal{N}$-extended deformed super-$\mathfrak{bms}\mathrm{_{3}}$\cite{Lodato:2016alv,Banerjee:2017gzj,Fuentealba:2017fck}. Another aspect that deserves further investigation is the supersymmetric extension of the Maxwell CS gravity with non-vanishing torsion (or Maxwellian Teleparallel gravity) \cite{Adami:2020xkm}, and the approach of the corresponding asymptotic symmetries, extending in this manner the asymptotic analysis done in \cite{Adami:2020xkm} to the supersymmetric case. We expect to find an infinite-dimensional lift of the Maxwellian teleparallel superalgebra as the asymptotic symmetry algebra (work in progress).

%%%%%%%%%%%%%%%%%%%%%%%%%%%%%%%%%%%%%%%%%%%%%%%%%%%%%%%%%%%%%%%%%%%%%%%%%%%%%%%%%%%%%%%%%%%%%

\section*{Acknowledgment}
This work was funded by the National Agency for Research and Development ANID - SIA grant No. SA77210097 and FONDECYT grant 11220486. The authors would like to thank to P. Concha for enlightening discussions and comments. ER would also like to thank the Direcci\'on de Investigaci\'on and Vice-rector\'ia de Investigaci\'on of the Universidad Cat\'olica de la Sant\'isima Concepci\'on, Chile, for their constant support. J.M. has been supported by the MCI, AEI, FEDER (UE)  grants PID2021-125700NB-C21 (“Gravity, Supergravity and Superstrings”(GRASS)) and IFT Centro de Excelencia Severo Ochoa CEX2020-001007-S.

%\section*{Appendix} \label{app}
%%%%%%%%%%%%%%%%%%%%%%%%%%%%%%%%%%%%%%%%%%%%%%%%%%%%%%%%%%%%%%%%%%%%%%%%%%%%%%%%%%%%%%%%%%%%%%%%%%%%

\bibliographystyle{fullsort}
 
\bibliography{Minimal_Maxwell}

\end{document}